\documentclass[aps,pra,secnumarabic,twocolumn,
groupedaddress,showpacs,amsmath,amssymb]{revtex4}%
\usepackage{amsmath}
\usepackage{amsfonts}
\usepackage{amssymb}
\usepackage{graphicx}
\usepackage{bm}%
\begin{document}
\title{Relaxation of the Bose-condensate oscillations in the mesoscopic system at T=0}
\author{Yu.\, Kagan}
\email{[]kagan@kurm.polyn.kiae.su}
\author{L.\, A.\, Maksimov}
\affiliation{Kurchatov Institute, Kurchatov Sq. 1, Moscow 123182, Russia}

\begin{abstract}
The general system is given of nonlinear equations describing dissipationless
evolution of the oscillating Bose-condensate. The relaxation of transverse
oscillations of the condensate in a trap of the cylindric symmetry is
considered. The evolution occurs due to parametric resonance coupling the
transverse mode with the longitudinal ones. The nonlinear rescattering in the
subsystem of discrete longitudinal modes results in suppression of the return
of energy, yielding dissipationless nonmonotonic relaxation of transverse
oscillations in the condensate.

\end{abstract}

\maketitle

The discovery of Bose-condensation in ultracold gases has yielded a unique
possibility for the study of evolution of the coherent properties in the
macroscopic system isolated from environment. One of most interesting aspects
here is connected with clarifying and analyzing the nature of the damping of
coherent oscillations in the condensate. So far, the theoretical and
experimental investigation of the problem has been connected, in fact, with
consideration of the damping at finite temperatures (see detailed bibliography
in \cite{1}). In this case the ensemble of normal excitations plays
effectively a role of the interior thermostat. The interaction with thermal
excitations results in damping the condensate oscillations with the transfer
of energy into the subsystem of normal excitations.

A question about the internal mechanism of damping at $T=0$ represents a
special interest in this problem. In the work of authors \cite{2} the
existence of such mechanism is shown on the example dealing with the damping
of the radial coherent oscillations in the condensate with the cylindric
symmetry in the transverse parabolic potential. It is found that the damping
arises from a special parametric resonance resulting in the transfer of energy
into the subsystem of longitudinal modes. The parametric resonance is due to
oscillations of the sound velocity as a result of transverse oscillations of
the condensate. It should be emphasized that the results obtained in \cite{2}
can be used only for describing the initial period of damping.

To describe the damping at the large times in the isolated system of finite
sizes with the discrete system of energy levels, it is necessary to know
additionally the temporal evolution in the subsystem of transverse modes and
involve variation of the parametric resonance with reducing the amplitude of
transverse oscillations. This aspect is a significant difference from the
conventional picture of parametric resonance. The choice of geometry of the
system is determined by the fact that, for an arbitrary variation of the
frequency in a two-dimensional parabolic potential of the circular symmetry
and arbitrary magnitudes of parameters, as is strictly found in \cite{3} (also
\cite{4}), two-dimensional oscillations do not damp.

Recently, Paris group \cite{1} has published the results of the experimental
study on the damping of transverse modes in the condensate as a whole
(breathing mode, BM) in the elongated trap with the azimuthal symmetry for
extremely low temperature of 40nK. At the small magnitude of the BM amplitude
the authors observe the record slow damping. For larger magnitude of the
initial BM amplitude, the picture changes drastically. After the fall within a
limited time interval the reverse transfer of energy with the growth of the BM
amplitude starts. The normal behavior of damping recovers only a noticeable
time interval later.

The theoretical analysis with explanation of the anormal picture is given in
\cite{5}. It is found that the involvement of nonlinear coupling between
longitudinal and transverse modes is essential. However, the relaxation of
longitudinal excitations was considered phenomenologically in essence.

In fact, for the discrete energy spectrum inherent in the closed mesoscopic
system, the irreversible damping does not appear. It can be spoken only about
the nonstationary redistribution of energy in the subsystem of longitudinal modes.

In the present work we have obtained a full set of nonlinear equations
describing the temporal evolution of the oscillating condensate in the lack of
the irreversible channels of dissipation. Our starting point is the
Gross-Pitaevskii equation. The condensate of the cylindrical symmetry is
considered in a transverse parabolic potential with longitudinal size $2L\gg
R$ where $R$ is the static radius of the condensate. The analysis of the
solution demonstrates an essential dependence of the damping of radial
oscillations upon the character of evolution of longitudinal modes.
Redistribution of the energy transferred into this subsystem results in the
chaotic nonstationary picture of filling the discrete levels and
simultaneously in the significant reduction for the amplitudes of active
longitudinal modes excited directly as a result of parametric resonance. This
reduction is equivalent effectively to relaxation but occurs in the lack of
real dissipation.

We will consider the dynamics of Bose-condensate in a rarified gas at $T=0$
within the framework of the nonlinear Gross-Pitaevskii equation for condensate
wavefunction $\Psi$
\begin{multline}
i\hbar\frac{\partial\Psi\left(  \vec{r},t\right)  }{\partial t}=[-\frac
{\hbar^{2}}{2m}\nabla^{2}+V\left(  \vec{r}\right)  ]\Psi\left(  \vec
{r},t\right) \\
+U_{0}\Psi^{\ast}\left(  \vec{r},t\right)  \Psi^{2}\left(  \vec{r},t\right)  .
\label{1}%
\end{multline}
Here $U_{0}=4\pi\hbar^{2}a/m$ is the vertex of the local interaction between
particles and $a$ is the scattering length. The system is assumed to be closed
and in the course of evolution the number of particles $N$ and total energy
$E$ conserve. We restrict ourselves by the case of repulsion between particles
($a>0$). Let us represent the wavefunction in the form
\[
\Psi\left(  \vec{r},t\right)  =\sqrt{n\left(  \vec{r},t\right)  }%
e^{i\Phi\left(  \vec{r},t\right)  }%
\]
The imaginary part of Eq. (\ref{1}) yields directly the equation of continuity
for the condensate
\begin{align}
\frac{\partial n_{1}\left(  \vec{r},t\right)  }{\partial t}+\nabla
\lbrack(n_{0}\left(  \vec{r}\right)  +n_{1}\left(  \vec{r},t\right)  ]\vec
{v}\left(  \vec{r},t\right)   &  =0,\label{14}\\
\vec{v}\left(  \vec{r},t\right)   &  =\frac{\hbar}{m}\nabla\Phi\left(  \vec
{r},t\right)  .\nonumber
\end{align}
In (\ref{14}) we introduced the fraction of density $n_{1}\left(  \vec
{r},t\right)  $ depending explicitly on the time and represented the total
density as $n\left(  \vec{r},t\right)  =n_{0}\left(  \vec{r}\right)
+n_{1}\left(  \vec{r},t\right)  $.

The real part of Eq. (\ref{1}) yields an equation for the phase
\begin{equation}
-\hbar\frac{\partial\Phi}{\partial t}=\frac{\hbar^{2}}{2m}(\nabla\Phi
)^{2}+U_{0}n+V\left(  \vec{r}\right)  -\frac{\hbar^{2}}{2m\sqrt{n}}\nabla
^{2}\sqrt{n}. \label{15}%
\end{equation}
Below we consider a gas of the sufficiently high density so that the
correlation length $\xi=1/\sqrt{mn_{0}U_{0}}$ is small compared with all sizes
of the system. It proves to be that only the long wave condensate excitations
with the wavelengths larger than $\xi$ are involved into the processes
considered. In these conditions corresponding to the Thomas-Fermi
approximation the last term in (\ref{15}) can be neglected, e.g., \cite{6}.
For the stationary case when $\vec{v}=0$, the phase of the condensate
wavefunction has the familiar value $\Phi_{0}=-\mu_{0}t/\hbar$ where $\mu_{0}$
is the chemical potential. Then from equation (\ref{15}) one has
\begin{equation}
n_{0}(\vec{r})=\frac{1}{U_{0}}\left(  \mu_{0}-V(\vec{r})\right)  \label{16}%
\end{equation}
Keeping notation $\Phi$ for the phase connected with the condensate dynamics,
we find
\begin{equation}
-\hbar\frac{\partial\Phi}{\partial t}=\frac{\hbar^{2}}{2m}(\nabla\Phi
)^{2}+U_{0}n_{1} \label{17}%
\end{equation}
A set of nonlinear equations (\ref{14}) and (\ref{17}) describes not only
excitations of quantum fluid but also their interaction. We will strictly take
this interaction into account, not involving the perturbation theory. Note
that the external potential enters equations (\ref{14}) and (\ref{17}) only
implicitly via the static distribution of density (\ref{16}).

Within the linear approximation the system of equations (\ref{14}) and
(\ref{17}) reduces to the equation of harmonic oscillations
\[
\frac{\partial^{2}n_{1}}{\partial t^{2}}=c_{0}^{2}\nabla\left(  f_{0}\nabla
n_{1}\right)  .
\]
Here
\begin{equation}
f_{0}(\vec{r})=\frac{n_{0}(\vec{r})}{n_{00}},\ n_{00}=n_{0}(0),\ c_{0}
^{2}=\frac{U_{0}n_{00}}{m}=\frac{\mu_{0}}{m}. \label{017}%
\end{equation}
Let us introduce the whole orthonormalized system of eigenfuctions of the
harmonic problem $\left\{  \chi_{s}(\vec{r})\right\}  $ being solution of the
equation
\begin{equation}
\Omega_{k}^{2}\chi_{s}(\vec{r})+c_{0}^{2}\nabla\left(  f_{0}(\vec{r}%
)\nabla\chi_{s}(\vec{r})\right)  =0. \label{00}%
\end{equation}
Let us expand $n_{1}$ and $\Phi$\ in the whole system of eigenfunctions
$\chi_{s}(\vec{r})$
\begin{equation}
n_{1}(t,\vec{r})= {\displaystyle\sum\limits_{s}} c_{s}(t)\chi_{s}(\vec
{r}),\ \ \ \Phi={\displaystyle\sum\limits_{s}} a_{s}(t)\chi_{s}(\vec{r})
\label{72}%
\end{equation}
Inserting these expansions into Eqs. (\ref{14}) and (\ref{17}), one finds
\begin{equation}%
\begin{array}
[c]{c}%
\frac{\partial}{\partial t}c_{s}-\frac{\hbar\Omega_{s}^{2}}{U_{0}} a_{s}%
=\frac{\hbar}{m}{\displaystyle\sum\limits_{s_{1}s_{2}}} c_{s_{1}}a_{s_{2}%
}\left\langle \chi_{s_{1}}\left(  \nabla\chi_{s}^{\ast}\right)  \left(
\nabla\chi_{s_{2}}\right)  \right\rangle \\
\frac{\partial}{\partial t}a_{p}+\frac{U_{0}}{\hbar}c_{p}=-\frac{\hbar}{2m}
{\displaystyle\sum\limits_{s_{1}s_{2}}} a_{s_{1}}a_{s_{2}}\left\langle
\chi_{s}^{\ast}\left(  \nabla\chi_{s_{1} }\right)  \left(  \nabla\chi_{s_{2}%
}\right)  \right\rangle
\end{array}
\label{04}%
\end{equation}
Here $\left\langle ...\right\rangle =\int d^{3}r(...)$. For the right-hand
side of the first equation, we employ the transformation
\[
\left\langle \chi_{s}^{\ast}\nabla(\chi_{s_{1}}\left(  \nabla\chi_{s_{2}
}\right)  )\right\rangle =-\left\langle \chi_{s_{1}}\left(  \nabla\chi
_{s}^{\ast}\right)  \left(  \nabla\chi_{s_{2}}\right)  \right\rangle
\]
Let us introduce new variables
\[
a_{s}=\left(  \frac{U_{0}}{\hbar\Omega_{s}}\right)  ^{1/2}(y_{s}+y_{-s}^{\ast
}),\ \ \ c_{s}=i\left(  \frac{\hbar\Omega_{s}}{U_{0}}\right)  ^{1/2}
(y_{s}-y_{-s}^{\ast})
\]
For the modes described by real eigenfunctions $\chi_{s}$, coefficients
$a_{s}$, $b_{s}$ are real and $y_{-s}\equiv y_{s}$.

For these variables, Eqs. (\ref{04}) can be represented as
\begin{multline}
\frac{\partial}{\partial t}y_{s}+i\Omega_{s}y_{s}= {\displaystyle\sum
\limits_{s_{1}s_{2}}} M_{ss_{1}s_{2}}(y_{s_{1}}-y_{-s_{1}}^{\ast})(y_{s_{2}%
}+y_{-s_{2}}^{\ast})\\
-\frac{1}{2} {\displaystyle\sum\limits_{s_{1}s_{2}}} K_{ss_{1}s_{2}}(y_{s_{1}%
}+y_{-s_{1}}^{\ast})(y_{s_{2}}+y_{-s_{2}}^{\ast}) \label{014}%
\end{multline}
Here
\begin{equation}%
\begin{array}
[c]{c}%
M_{ss_{1}s_{2}}=\frac{1}{2m}\left(  \frac{\hbar U_{0}\Omega_{s_{1}}}
{\Omega_{s}\Omega_{s_{2}}}\right)  ^{1/2}\left\langle \left(  \nabla\chi
_{s}^{\ast}\right)  \chi_{s_{1}}\left(  \nabla\chi_{s_{2}}\right)
\right\rangle ,\\
K_{ss_{1}s_{2}}=\frac{1}{2m}\left(  \frac{\hbar U_{0}\Omega_{s}}{\Omega
_{s_{1}}\Omega_{s_{2}}}\right)  ^{1/2}\left\langle \chi_{s}^{\ast}\left(
\nabla\chi_{s_{1}}\right)  \left(  \nabla\chi_{s_{2}}\right)  \right\rangle .
\end{array}
\label{016}%
\end{equation}
Let at the initial time moment a transverse condensate oscillation alone be
excited as a whole with conservation of the cylindrical symmetry ("breathing
mode" and hereafter index $\perp$). A set of equations (\ref{014}) can be
represented in this case as
\begin{equation}%
\begin{array}
[c]{c}%
\frac{\partial}{\partial t}y_{\perp}+i\Omega_{\perp}y_{\perp}=
{\displaystyle\sum\limits_{k}}
M_{\perp k-k}(y_{k}y_{-k}-y_{k}^{\ast}y_{-k}^{\ast})\\
-\frac{1}{2}%
{\displaystyle\sum\limits_{k}}
K_{\perp k-k}(y_{k}+y_{-k}^{\ast})(y_{-k}+y_{k}^{\ast}),\\
\frac{\partial}{\partial t}y_{k}+i\Omega_{k}y_{k}=-2M_{kk\perp}y_{-k}^{\ast
}(y_{\perp}+y_{\perp}^{\ast})\\
+K_{\perp k-k}(y_{\perp}-y_{\perp}^{\ast})(y_{k}+y_{-k}^{\ast})+I_{k}.
\end{array}
\label{71}%
\end{equation}
In the second equation we use equalities $M_{k\perp k}=K_{\perp k-k}%
,\ K_{kk\perp}=M_{kk\perp}$. The last term on the right-hand side determines
the evolution in the longitudinal subsystem of excitations
\begin{multline}
I_{k}=
{\displaystyle\sum\limits_{k_{1}k_{2}}}
M_{kk_{1}k_{2}}(y_{k_{1}}-y_{-k_{1}}^{\ast})(y_{k_{2}}+y_{-k_{2}}^{\ast})\\
-\frac{1}{2}%
{\displaystyle\sum\limits_{k_{1}k_{2}}}
K_{kk_{1}k_{2}}(y_{k_{1}}+y_{-k_{1}}^{\ast})(y_{k_{2}}+y_{-k_{2}}^{\ast}).
\label{721}%
\end{multline}
The eigenfunction of the breathing mode is found from the solution of equation
(\ref{00}) involving that within the Thomas-Fermi approximation $f_{0}%
=1-r^{2}/R^{2}$ where $R^{2}=2c_{0}^{2}/\omega_{\perp}^{2}$, $\omega_{\perp}$
being the frequency of a parabolic trap.
\begin{equation}
\chi_{\perp}=\sqrt{\frac{3}{V}}(1-\frac{2r^{2}}{R^{2}}),\ \ \Omega_{\perp
}=2\omega_{\perp}, \label{018}%
\end{equation}
Quantity $k$ determines the value of wave vector for longitudinal excitations,
running discrete values due to finite sizes of the system in the $z$
direction. Following work \cite{7}, it is easily to find the eigenfunctions
and eigenvalues for the longwave longitudinal modes
\begin{equation}%
\begin{array}
[c]{c}%
\chi_{k}=\sqrt{\frac{1}{V}}e^{ikz}[1-\frac{\left(  kR\right)  ^{2}}
{16}(1-\frac{2r^{2}}{R^{2}})],\\
\Omega_{k}=\bar{c}k(1-\frac{\left(  kR\right)  ^{2}}{96}),\ \bar{c}%
=\frac{c_{0}}{\sqrt{2}}.
\end{array}
\label{019}%
\end{equation}
Using (\ref{018}) and (\ref{019}), one can straightforwardly calculate matrix
elements (\ref{016}) entering in (\ref{71})
\begin{equation}%
\begin{array}
[c]{c}%
M_{\perp k-k}=M_{kk\perp}=-\eta\frac{\sqrt{6}}{4}\left(  \frac{\Omega_{k}%
}{\omega_{\perp}}\right)  ^{2},\\
K_{\perp k-k}=-\eta\frac{5\sqrt{6}}{24}\left(  \frac{\Omega_{k}}{\omega
_{\perp}}\right)  ^{3},\ \eta=\omega_{\perp}\left(  \frac{\hbar\omega_{\perp}%
}{Vn_{00}\mu_{0}}\right)  ^{1/2}%
\end{array}
\label{21}%
\end{equation}
and in (\ref{721})
\begin{equation}%
\begin{array}
[c]{c}%
M_{kk_{1}k_{2}}=\eta\left(  \frac{\Omega_{k}\Omega_{k_{1}}\Omega_{k_{2}}%
}{\omega_{\perp}^{3}}\right)  ^{1/2}\text{sign}(kk_{2})\delta_{k,k_{1}+k_{2}%
},\\
K_{kk_{1}k_{2}}=-\eta\left(  \frac{\Omega_{k}\Omega_{k_{1}}\Omega_{k_{2}}%
}{\omega_{\perp}^{3}}\right)  ^{1/2}\text{sign}(k_{1}k_{2})\delta
_{k,k_{1}+k_{2}}%
\end{array}
\label{211}%
\end{equation}
Equations (\ref{71}) and (\ref{721}) with matrix elements (\ref{21}) and
(\ref{211}) describe fully the relaxation of the transverse condensate
oscillations in the mesoscopic system in the lack of dissipative channels for evolution.

We assume that the initial amplitude of breathing mode is relatively small,
i.e., $\left\vert \delta R/R\right\vert _{0}\ll1$. This restricts a scale of
the nonlinear interaction between modes. From the other side the energy
interval in which the discrete longitudinal modes are connected effectively
with the breathing mode proves to be small compared with $\omega_{\perp}$.
Under conditions we can simplify a set of equations, restricting ourselves
with the quasi resonance approximation. Within the framework of the
approximation in equations (\ref{71}) and (\ref{721}) we can retain only the
terms for which the following inequality is valid
\begin{equation}
\Delta\Omega=\left\vert \Omega_{s}\pm\Omega_{s_{1}}\pm\Omega_{s_{2}
}\right\vert \ll\Omega_{s_{i}} \label{015}%
\end{equation}
Let us rewrite equations (\ref{71}) and (\ref{721}) within this approximation,
substituting ratios $\bar{y}_{s}=y_{s}(t)/\left\vert y_{\perp}(0)\right\vert $
for amplitudes $y_{s}$. The initial amplitude of breathing mode $y_{\perp}(0)$
can be found from comparing the vibrational energy of the condensate at the
initial time moment $E_{vib}=\frac{4}{3}\mu_{0}N\left\vert \delta
R/R\right\vert _{0}^{2}$ with the energy of transverse mode $2\hbar
\Omega_{\perp}\left\vert y_{\perp}(0)\right\vert ^{2}$. Hence
\[
\left\vert y_{\perp}(0)\right\vert =\left(  \frac{2\mu_{0}N}{3\hbar
\Omega_{\perp}}\right)  ^{1/2}\left\vert \frac{\delta R}{R}\right\vert _{0}
\]
After involvement of these notations equations (\ref{71}) and (\ref{721}) go
over into
\[
\frac{\partial}{\partial t}\bar{y}_{\perp}+i\Omega_{\perp}\bar{y}_{\perp
}=-\alpha\underset{k>0}{\sum}^{\prime}\bar{y}_{k}\bar{y}_{-k},
\]
A dash in a sum of the first equation means summation over "active modes"
alone, interacting directly with the BM according to restriction (\ref{015})
\begin{multline}
\frac{\partial}{\partial t}\bar{y}_{k}+i\Omega_{k}\bar{y}_{k}=\alpha\bar
{y}_{-k}^{\ast}\bar{y}_{\perp}\\
+\alpha^{\prime} {\displaystyle\sum\limits_{k>k_{1}>0}} \left(  \frac
{\Omega_{k}\Omega_{k_{1}}\Omega_{k-k_{1}}}{\omega_{\perp}^{3} }\right)
^{1/2}\bar{y}_{k_{1}}\bar{y}_{k-k_{1}}\\
-2\alpha^{\prime} {\displaystyle\sum\limits_{k_{1}>0}} \left(  \frac
{\Omega_{k}\Omega_{k_{1}}\Omega_{k+k_{1}}}{\omega_{\perp}^{3} }\right)
^{1/2}\bar{y}_{k_{1}}^{\ast}\bar{y}_{k+k_{1}},\ \ (k>0) \label{31}%
\end{multline}

For longitudinal modes which do not interact directly with the transverse
mode, one should omit the first term on the right-hand side of Eq. (\ref{31}).
In these equations
\begin{equation}%
\begin{array}
[c]{l}%
\alpha=\left\vert \left(  2M_{\perp k-k}-K_{\perp k-k}\right)  \right\vert
\left\vert y_{\perp}(0)\right\vert \approx0,3\omega_{\perp}\left\vert
\frac{\delta R}{R}\right\vert _{0},\\
\alpha^{\prime}\approx2\alpha.
\end{array}
\label{39}%
\end{equation}

Let us take that one has $\bar{y}_{k}(0)=\bar{y}_{-k}(0)$ for the trivially
degenerated states at the initial time moment. It follows from a set of
equations (\ref{31}) that this equality holds in the course of evolution. To
analyze, it is convenient to select the fast phase from complex variables
$\bar{y}_{s}$, representing them as
\[
\bar{y}_{s}=b_{s}\left(  t\right)  e^{-i(\Omega_{s}t+\varphi_{s}\left(
t\right)  )},
\]
where $b_{s}\left(  t\right)  $ and $\varphi_{s}\left(  t\right)  $ are the
real quantities varying relatively slow in time at $\alpha/\omega_{\perp}\ll
1$. Let us introduce notations
\begin{equation}
\gamma_{kk_{1}k_{2}}=(\Omega_{k}-\Omega_{k_{1}}-\Omega_{k_{2}})t+(\varphi
_{k}-\varphi_{k_{1}}-\varphi_{k_{2}}) \label{34}%
\end{equation}
Separating real and imaginary parts (\ref{31}), one finds
\begin{equation}
\frac{\partial}{\partial t}b_{\perp}=-\alpha\underset{k>0}{\sum}^{\prime}%
b_{k}^{2}\cos\gamma_{\perp kk},\ \frac{\partial}{\partial t}\varphi_{\perp
}=\alpha\underset{k>0}{\sum}^{\prime}\frac{b_{k}^{2}}{b_{\perp}}\sin
\gamma_{\perp kk}, \label{35}%
\end{equation}
\begin{equation}%
\begin{array}
[c]{c}%
\frac{\partial}{\partial t}b_{k}=\alpha b_{k}b_{\perp}\cos\gamma_{\perp
kk}+I_{k},\\
I_{k}=\alpha^{\prime}%
{\displaystyle\sum\limits_{k>k_{1}>0}}
\left(  \frac{\Omega_{k}\Omega_{k_{1}}\Omega_{k-k_{1}}}{\omega_{\perp}^{3}%
}\right)  ^{1/2}b_{k_{1}}b_{k-k_{1}}\cos\gamma_{kk_{1}k-k_{1}}\\
-2\alpha^{\prime}%
{\displaystyle\sum\limits_{k_{1}>0}}
\left(  \frac{\Omega_{k}\Omega_{k_{1}}\Omega_{k+k_{1}}}{\omega_{\perp}^{3}%
}\right)  ^{1/2}b_{k+k_{1}}b_{k_{1}}\cos\gamma_{k+k_{1}k_{1}k},
\end{array}
\label{36}%
\end{equation}%
\begin{equation}%
\begin{array}
[c]{c}%
\frac{\partial}{\partial t}\varphi_{k}=\alpha b_{\perp}\sin\gamma_{\perp
kk}+I_{k}^{\prime},\\
I_{k}^{\prime}=-\alpha^{\prime}%
{\displaystyle\sum\limits_{k>k_{1}>0}}
\left(  \frac{\Omega_{k}\Omega_{k_{1}}\Omega_{k-k_{1}}}{\omega_{\perp}^{3}%
}\right)  ^{1/2}\frac{b_{k_{1}}b_{k-k_{1}}}{b_{k}}\sin\gamma_{kk_{1}k-k_{1}}\\
-2\alpha^{\prime}%
{\displaystyle\sum\limits_{k_{1}>0}}
\left(  \frac{\Omega_{k}\Omega_{k_{1}}\Omega_{k+k_{1}}}{\omega_{\perp}^{3}%
}\right)  ^{1/2}\frac{b_{k+k_{1}}b_{k_{1}}}{b_{k}}\sin\gamma_{k+k_{1}k_{1}k}.
\end{array}
\label{37}%
\end{equation}

The analysis of the equations obtained allows us to find a scenario of
relaxation of the transverse condensate oscillations. At the initial time
moment the longitudinal modes are not excited and relation $b_{k}\left(
0\right)  \ll1$ is valid for them, while according to definition $b_{\perp
}\left(  0\right)  =1$. Thus at the initial time period the right-hand sides
of Eq.~ (\ref{35}) and terms $I_{k}$, $I_{k}^{\prime}$ quadratic in $b_{k}$
play no role and evolution, in essence, is governed by first terms on the
right-hand sides of equations (\ref{36}) and (\ref{37}) with constant
$b_{\perp}$. The joint solution demonstrates an exponential growth $b_{k}(t)$
provided
\begin{equation}
\left\vert \Delta\Omega\right\vert \equiv\left\vert \Omega_{k}-\frac{1}%
{2}\Omega_{\perp}\right\vert <\alpha\,. \label{38}%
\end{equation}
In particular, for $\left\vert \Delta\Omega\right\vert =0$, one has
$\gamma_{\perp kk}\simeq0$ and equation (\ref{36}) yields $b_{k}%
(t)=b_{k}\left(  0\right)  \exp\alpha t$. These results, involving requirement
of the finiteness for initial amplitude $b_{k}\left(  0\right)  $, are a
typical manifestation of parametric resonance (cp. \cite{2}). All longitudinal
modes within the energy interval of about $\alpha$ (\emph{active modes})
experience an exponential growth. For the length of cylinder $2L$, the spacing
between levels equals $\delta\Omega=\pi\bar{c}/L$. If $\delta\Omega<2\alpha$,
at least, one mode lies within this interval. The growth of $b_{k}(t)$ results
in reduction of $b_{\perp}(t)$ after some delay and transfer of energy to the
other parallel modes at the same time. For $L\gg R$, there is a large number
of such modes but a discrete character of energy spectrum results in the total
prohibition of irreversible processes. However, irregular character of the
transition amplitudes like $\left(  \frac{\Omega_{k}\Omega_{k_{1}}%
\Omega_{k_{2}}}{\omega_{\perp}^{3}}\right)  ^{1/2}\cos\gamma_{kk_{1}k_{2}}$ in
(\ref{36}) and (\ref{37}) in nonlinear terms $I_{k}$,$I_{k}^{\prime}$ leads to
chaotic evolution in the system of longitudinal modes. One should think that
in this case the return to the active longitudinal levels should significantly
be suppressed and a noticeable fraction of the transferred energy should
remain in inactive modes.%

\begin{figure}
[ptb]
\begin{center}
\includegraphics[
height=1.8645in,
width=3.1903in
]%
{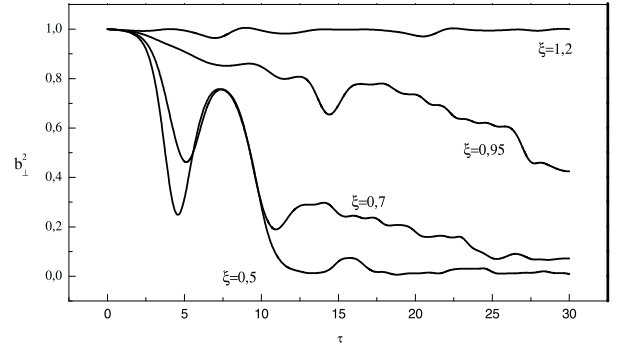}%
\caption{The behavior for the population of breathing mode as a function of
the time at fixed value $k_{0}=10\pi/L$ for $\xi=0.5,\ 0.7,\ 0.95$, and
$1.2$.}%
\label{Fig.1}%
\end{center}
\end{figure}
%

\begin{figure}
[ptb]
\begin{center}
\includegraphics[
height=1.8654in,
width=3.1946in
]%
{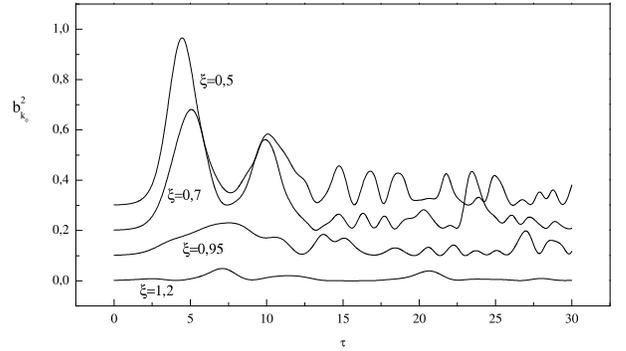}%
\caption{The behavior for the population of active mode $k_{0}$ as a function
of the time at the same values $\xi$. The plots are shifted with respect to
each other by $0.1$ along the vertical.}%
\label{Fig. 2}%
\end{center}
\end{figure}

The direct numerical simulation of system (\ref{35})-(\ref{37}) displays the
picture described. Let only single level $k_{0}$ lie within the energy
interval of about $\alpha$. We neglect weak dispersion of longitudinal modes
(\ref{019}). Then $\frac{\Omega_{k}\Omega_{k_{1}}\Omega_{k_{2}}}{\omega
_{\perp}^{3}}\cong\frac{kk_{1}k_{2}}{k_{0}}$ and in expression (\ref{34}) for
longitudinal modes one can omit the first term. Let us introduce dimensionless
time $\tau=\alpha t$ and take into account that $\alpha^{\prime}%
/\alpha=\text{const}$ (see, (\ref{39})). Then the evolution described depends,
practically, on ratio $\xi=\Delta\Omega/\alpha$ alone and, to weak extent, on
$k_{0}$ for $k_{0}L/\pi\gg1$. (Note that this is valid not only in the quasi
resonance approximation but also for the general system of equations
(\ref{71}).) In Fig.1 the dependence $b_{\perp}^{2}(\tau)$ $\equiv\left(
\delta R(\tau)/\delta R(0)\right)  ^{2}$ is given at various magnitudes of
parameter $\xi$ for fixed value $k_{0}$ (in units $\pi/L$). Here we involved
not only the discrete levels lying below the active level but also the
longitudinal levels above $k_{0}$ which start to be occupied at the next stage
after decay of state $\pm k_{0}$. The law of energy conservation, which system
(\ref{35})-(\ref{37}) satisfies, results in a weak population of the upper
levels in the course of evolution. In calculations we restricted ourselves by
the same number of levels above and below $\Omega_{k_{0}}$. For value
$\xi=1,\ 2$, the transverse oscillations does not decay at all. This is
obviously seen from Fig.2 in which the behavior of population $b_{k_{0}}^{2}$
of level $k_{0}$ on $\tau$ is plotted for the same values of parameter $\xi$.
For given value $\xi$, magnitude $b_{k_{0}}^{2}(\tau)$ remains close to zero
for all $\tau$. The statement that the damping of oscillations at $T=0$ should
be absent at all with violation of the conditions for appearance of parametric
resonance (\ref{38}) finds thus a direct confirmation. The curves
corresponding to $\xi=0.7$ and $0.5$, on the contrary, demonstrate an
origination of damping with the nontrivial character (see, Figs.~ 1 and 2).
Reduction $b_{\perp}^{2}(\tau)$ and growth $b_{k_{0}}^{2}(\tau)$ are followed
by the reverse transfer of energy to the BM and the decrease of the population
of the active mode. Only later, when the redistribution of energy over the
other longitudinal modes becomes noticeable, there arises a traditional
monotonic behavior of relaxation. The energy, as direct calculations show, is
spread over all longitudinal modes, their population experiencing chaotic
evolution. At this stage, in spite of the lack of irreversible processes, the
character of interference due to dispersion of dynamic phases prevents from
any noticeable growth of amplitude $b_{k_{0} }(\tau)$ and, thus, from reverse
transfer of energy to the breathing mode. Then the evolution looks like
relaxation of the transverse condensate oscillation with concentration of
larger fraction of energy in the subsystem of longitudinal modes.

For the first time, nonmonotonic character of the damping of transverse
breathing mode, analogous to that shown in Fig.1, is observed experimentally
in \cite{1}. Large magnitude $\left\vert \delta R/R\right\vert _{0}$ provides
condition $\xi<1$. It should be noted that the theoretical results obtained
are universal to essential extent. Thus the displayed picture of the effective
relaxation with its anomalously nonmonotonic behavior has a general character
at $T=0$.

However, quantitative comparison with the experimental results \cite{1} is
difficult, in the first turn, due to the finiteness of temperature. Though
$T<\mu_{0}$ in the experiment, but $T>\hbar\omega_{\perp}$ and the
longitudinal levels prove to be temperature-occupied at the initial time
moment. (Note that with the introduction of random dispersion of phases
$\varphi_{k}$ at the initial time moment the qualitative picture of evolution
holds but quantitative picture changes. The slow damping close to monotonic
and observed in \cite{1} at sufficiently smaller value $\left\vert \delta
R/R\right\vert _{0}$ has analog in an isolated system at $T=0$ only for $\xi$
close to unity (see, Fig.1 and $\xi=0.95$). However, one cannot exclude that
such damping is associated with the dissipation processes due to external
factors. For sufficiently large magnitude of ratio $\left\vert \delta
R/R\right\vert _{0}$, these processes cannot play a role at most interesting
stages of nonmonotonic relaxation.

The authors are grateful to D.L. Kovrizhin for help in numerical simulations.

The present work is supported by RFBR, INTAS and NWO (Netherlands).

\end{document}